\documentstyle[12pt,preprint,aps]{revtex}

\begin{document}
%\draft

\title{Analytical approach to viscous fingering in a cylindrical Hele-Shaw 
cell}

\author{Jos\'e A. Miranda\footnote{e-mail:jme@lftc.ufpe.br}}
\address{Laborat\'{o}rio de F\'{\i}sica Te\'{o}rica e Computacional,
Departamento de F\'{\i}sica,\\ Universidade Federal de Pernambuco, 
Recife, PE  50670-901 Brazil}
%\date{\today}
\maketitle

\begin{abstract} 
We report analytical results for the development of the viscous fingering 
instability in a cylindrical Hele-Shaw cell of radius $a$ and thickness $b$. 
We derive a generalized version of Darcy's law in such cylindrical background, 
and find it recovers the usual Darcy's law for flow in flat, rectangular 
cells, with corrections of higher order in $b/a$. We focus our interest 
on the influence of cell's radius of curvature on the instability 
characteristics. Linear and slightly nonlinear flow regimes are studied 
through a mode-coupling analysis. Our analytical results reveal that linear 
growth rates and finger competition are inhibited for increasingly larger 
radius of curvature. The absence of tip-splitting events in cylindrical cells 
is also discussed.
\end{abstract}


\section{INTRODUCTION}
\label{intro}
When a fluid is pushed by a less viscous one in a narrow 
space between two parallel plates (a device known as Hele-Shaw cell), 
the Saffman-Taylor instability arises~\cite{Saf}. This hydrodynamic
instability results in the complex evolution of the moving interface between 
the fluids, producing a wide range of patterns~\cite{Rev}. Since the 
pioneering work of Saffman and Taylor~\cite{Saf}, these visually striking, 
viscous fingering patterns have been extensively studied, both 
theoretically and experimentally~\cite{Rev}.

One noteworthy distinction between experimental and theoretical work on the 
Saffman-Taylor problem is that the measurements usually reveal disturbances 
associated with friction at the sidewalls of the Hele-Shaw cell, 
while most theoretical studies sidestep this problem by assuming periodic 
boundary conditions. In principle, rigid-wall boundary conditions would be 
closer to an experimental realization of the problem. However, the use of 
periodic boundary conditions, as opposed to rigid-wall ones in theoretical 
investigations of the Saffman-Taylor problem has been a matter of
interest and debate during the last few decades~\cite{Rev,Tho,Try1,Zhao,Cas}.

In the early 1990's an attempt to measure the flow in a cell which 
approximates the periodic boundary conditions of the theory has been 
performed by Zhao and Maher~\cite{Zhao}. Their interesting experimental work 
considered gravity-driven viscous flow within a {\it cylindrical} 
Hele-Shaw cell (two coaxial cylinders separated by a small gap) of large
radius of curvature. To drive the system gravitationally they allowed the 
fluids to form a stable flat interface, and then invert the cell to put the 
denser fluid on the top. The fluids used in~\cite{Zhao} had almost the 
same viscosity (low viscosity contrast), such that 
the process of finger competition leads to a nearly up-down symmetric 
interface. The main goal in Ref.~\cite{Zhao} was to directly compare
viscous fingering flow in a cylindrical cell with flow in a conventional, flat,
rectangular cell, and verify if the rigid sidewalls could be viewed as having 
been replaced with periodic boundary conditions. 

Comparison of the pattern evolution observed in the cylindrical cell 
experiment~\cite{Zhao} with similar measurements in rectangular cells 
with sidewalls~\cite{DiF1,DiF2}, shows that periodic boundary conditions yield 
{\it few differences} of results. Moreover, the cylindrical cell patterns 
also look very similar to computer simulation results obtained by Tryggvason 
and Aref~\cite{Try1}, who used periodic boundary conditions in a flat 
rectangular cell. 

The experimental results presented in~\cite{Zhao} suggest 
that, for large radius of curvature and low viscosity contrast, there are no 
significant changes made in the statistical properties of 
the Saffman-Taylor flow by friction at the boundaries. In other words, it 
seems that the boundaries affect the dynamics in much the same 
way whether they are provided by periodic boundary conditions or by the 
presence of physical sidewalls. Although the majority of flow features
survive the elimination of sidewall friction, the authors in Ref.~\cite{Zhao}
were not able to perform a meaningful test of growth rates in the linear
regime due to experimental disturbances other than rigid-wall ones, related to 
the inversion of the cell. It is worth noting that sidewall effects are more 
prominent in cases with higher viscosity contrast~\cite{Park,Maher,Max}.

Despite the relevance and simplicity of the experimental work carried out in 
Ref.~\cite{Zhao}, a theoretical analysis of the viscous fingering instability 
in a cylindrical Hele-Shaw cell is lacking in the literature. A theoretical
study of flow in cylindrical cells, could yield new insigth into the
possible causes that may contribute to the few differences of results detected 
in~\cite{Zhao}. In this work, we examine flow in cylindrical cells 
analytically. We consider the general case of arbitrary viscosity contrast and 
cell radius. In section~\ref{darcylinder}, 
we derive a generalized version of Darcy's law 
suitable to describe flow in cylindrical cells. This Darcy's law introduces a 
correction factor, which depends on the ratio between the cell's thickness and 
radius. This result enables us to express the differences of behavior 
between flow in rectangular and cylindrical cells in quantitative terms. In 
section~\ref{discuss} we investigate the 
consequences of such differences in the flow dynamics by performing a 
mode-coupling analysis of the problem. We examine both linear and slightly 
nonlinear stages of evolution, and explicity show how linear growth rates and
the dynamical process of finger competition are influenced by
cylindrical geometry. The absence of finger tip-splitting in cylindrical cells 
is briefly discussed. Section~\ref{final} presents our final remarks.

\section{DARCY'S LAW IN CYLINDRICAL ENCLOSURES}
\label{darcylinder}
In this section we present the physical system of interest and derive a 
generalized Darcy's law which is suited to bringing out the geometrical 
aspects related to viscous fluid flow in cylindrical passages.
Consider two immiscible, incompressible, viscous fluids, flowing in a
narrow gap of thickness $b$, between two long coaxial, thin right circular 
cylinders (cylindrical Hele-Shaw cell). The radius of curvature of the 
cylindrical cell is $a$ (see Fig. 1). Denote the densities 
and viscosities of the lower and upper fluids, 
respectively as $\rho_{1}$, $\eta_{1}$ and $\rho_{2}$, $\eta_{2}$. 
The flows in fluids $1$ and $2$ are assumed to be irrotational, and between 
them there exists a surface tension $\sigma$. 
The acceleration of gravity is represented by ${\bf g}$, and points 
downward along the direction of cylinders' common axis.

In order to derive a generalized version of Darcy's law, adjusted to 
describe flow in such confined cylindrical environment, it suffices to
focus on a single fluid. The starting point of our calculation is a 
coordinate-free 
representation of the continuity equation for an incompressible fluid
\begin{equation}
\label{cont}
{\nabla} \cdot {\bf u} = 0,
\end{equation}
and the Navier-Stokes equation
\begin{equation}
\label{NS-CF}
\rho \left [ {{\partial {\bf u}}\over{\partial t}} 
+ ({\bf u} {\cdot \nabla}) {\bf u} \right ]
= - { \nabla} p + \eta { \nabla}^{2} {\bf u} + \rho {\bf g},
\end{equation}
where ${\bf u}$ denotes the three-dimensional fluid velocity and $p$ 
is the hydrodynamic pressure. 

We consider cylindrical coordinates $(r,\varphi, z)$, where the $z$ axis 
coincides with that of the two cylinders. Specializing to the case of 
flow in the $z$ direction ${\bf u}_{z}(r,z)=u_{z}(r,z) {\bf \hat{\rm z}}$, the 
continuity equation~(\ref{cont}) leads to $\partial (r ~u_{z})/ \partial z= 
0$. Consequently, the velocity can be written is a function of $r$ only
\begin{equation}
\label{general}
{\bf u}_{z}=u(r) ~\hat{z},
\end{equation}
where $\hat{z}$ denotes the unit vector along the $z$ axis.

Following the standard approach in Hele-Shaw problems, we restrict 
our attention to small velocity flows of viscous fluids, and 
neglect the inertial terms on the left-hand side of Eq.~(\ref{NS-CF}).
Under such circumstances, we use the solution~(\ref{general}) to 
rewrite the Navier-Stokes equation as
\begin{equation}
\label{NS-SC}
{{\partial p}\over{\partial z}} - \rho g
= \frac{\eta}{r} ~{{\partial }\over{\partial r}} \left [  r {{\partial 
~u(r)}\over{\partial r}}    \right ].
\end{equation}
Since the left-hand side is a function of $z$, and the right-hand side 
involves only $r$, Eq.~(\ref{NS-SC}) can be satisfied only if 
each side is equal to a constant of common value $B$. Imposing the no-slip 
boundary conditions at the solid cylindrical 
shells $u(a)=u(a+b)=0$, we find the solution of the radial
equation
\begin{equation}
\label{uofr}
u(r)={{B}\over{4 \eta}} \left [ r^{2} - C \log{\left ( \frac{r}{a} \right )} - 
a^{2} \right ],
\end{equation}
where $C=[(a + b)^{2} - a^2]/\log{(1 + b/a)}$. In contrast to flow in usual 
flat, 
rectangular cells, observe that the velocity profile Eq.~(\ref{uofr}) is not 
rigorously parabolic (or, Poiseuille-like) due to the presence of a 
logarithmic term. The profile is very close to a parabola for $r/a \ll 1$, but 
deviates from parabolic shape for larger values of $r/a$.

Averaging the three-dimensional velocity ${\bf u}$ with respect to the 
transverse, radial direction, defining $v \equiv (1/b)~\int_{a}^{a + b} u(r) 
dr$, 
leads to the equation for the mean flow velocity of the fluid
\begin{equation}
\label{ave}
v= - {{b^2 ~F(b/a)}\over{12\eta}} \left [{{\partial 
p}\over{\partial z}} - \rho g \right ],
\end{equation}
where 
\begin{equation}
\label{eff}
F(b/a)=3 \left ( \frac{a}{b} \right )^{3} 
\left \{ \frac{\frac{b}{a} \left [ 1 - \left (1 + \frac{b}{a} \right )^{2} 
\right ]}{\log{\left ( 1 + \frac{b}{a} \right)}} - \frac{2}{3} \left [1 - 
\left (1 + \frac{b}{a} \right )^{3} \right ] \right \}. 
\end{equation}
Equation~(\ref{ave}) is Darcy's law for flow in cylindrical Hele-Shaw cells. 
The function $F(b/a) \approx 1 ~+ ~(1/30) ~(b/a)^{2} ~+ ~ {\cal O} 
[(b/a)^{3}]$ ~introduces a 
correction factor which measures the deviation from the flat, rectangular case 
($a \rightarrow \infty$). Equation~(\ref{ave}) recovers the usual 
Darcy's law for flow in flat cells with corrections of higher 
order in $b/a$, introduced by $1 \le F(b/a) \le 2$ .

Darcy's law is the governing equation for Hele-Shaw-type flows~\cite{Rev}. 
Therefore, quantitative comparisons between dynamical behavior of flow in 
rectangular and cylindrical cells must take into account the corrections 
introduced by Eq.~(\ref{ave}), in addition to those caused by sidewall effects 
and cell inversion~\cite{Zhao}. In contrast to the unavoidable disturbances 
mentioned in Ref.~\cite{Zhao}, which are hard to quantify 
accurately, Eq.~(\ref{ave}) allows precise determination 
of intrinsic, purely geometrical effects.

Based on Eq.~(\ref{eff}) we can start understanding why the differences 
of results detected in Ref.~\cite{Zhao} were so small. In Ref.~\cite{Zhao} the 
authors used a cylindrical cell of radius of curvature $a= 18 ~{\rm mm}$ and 
thickness $b=1 ~{\rm mm}$, such that $b/a \approx  5.6 \times 10^{-2}$. From 
Eq.~(\ref{eff}) this means a small correction of 0.01 $\%$ with respect to the 
flat cell case ($a \rightarrow \infty$). In practical terms, we estimate that 
a ratio of roughly $b/a \approx 10^{-3}$ would be enough in order for 
curvature effects to be considered negligible.

\section{LINEAR AND SLIGHTLY NONLINEAR DYNAMICS}
\label{discuss}

In this section we investigate the consequences of the changes 
introduced by the generalized Darcy's law~(\ref{ave}) in both linear and 
weakly nonlinear stages of the interface evolution. We focus on two general 
questions: (i) on the linear level, what is the effect of Eq.~(\ref{ave}) on 
linear growth rates? (ii) concerning the onset of nonlinear effects,
how do finger competition, up-down interfacial symmetry, and finger 
tip-splitting  are influenced by
Eq.~(\ref{ave})?

To study these issues we derive a second order mode-coupling differential 
equation for the interface perturbation amplitudes. We express the
fluid-fluid interface as a Fourier series
\begin{equation}
\label{z}
\zeta(\varphi,t)=\sum_{n=-\infty}^{+\infty} \zeta_{n}(t) \exp{(i n \varphi)}, 
\end{equation}
where $\zeta_{n}(t)$ denotes the complex Fourier mode amplitudes 
and $n$=0, $\pm 1$, $\pm 2$, $...$ is the discrete azimuthal wave number.

We exploit the irrotational flow condition to define the velocity
potential ${\bf v}_{j}=-{\bf \nabla} \phi_{j}$ in fluids $j=1$ and $j=2$.  
Using the velocity potential, we evaluate Darcy's law~(\ref{ave}) for each of 
the fluids on the interface, subtract the resulting expressions from each 
other, and divide by the sum of the two fluids' viscosities to get the
equation of motion
\begin{equation}
\label{dimensionless2}
A \left ( \frac{\phi_{1}|_{{\zeta}} + \phi_{2}|_{{\zeta}}}{2} \right ) -  
\left ( \frac{\phi_{1}|_{{\zeta}} - \phi_{2}|_{{\zeta}}}{2} \right ) = F(b/a) 
~\left [U z + \kappa \right ]|_{\zeta}.
\end{equation}
Note the presence of the correction factor~(\ref{eff}) on the right-hand side 
of Eq.~(\ref{dimensionless2}). To obtain~(\ref{dimensionless2}) we used the 
pressure boundary condition 
$p_{2} - p_{1}=\sigma\kappa$ at the interface $z=\zeta$, where 
$\kappa=(1/a^{2})( \partial^2 \zeta/ \partial \varphi^{2} ) [ 1 + ( 
\partial\zeta/
\partial \varphi)^2]^{-3/2}$ is the interfacial curvature. The viscosity 
contrast A=$(\eta_{2} - \eta_{1})/(\eta_{2} + \eta_{1})$, and $U=b^{2} g 
~(\rho_{2} - \rho_{1})/[12(\eta_{1} + \eta_{2})]$ is a characteristic 
velocity. In Eq.~(\ref{dimensionless2}) we introduced dimensionless variables, 
scaling all lengths by the gap size $b$, and all velocities by 
$\sigma/12(\eta_1+\eta_2)$. From now on we work, unless otherwise stated, with 
the dimensionless version of Eq.~(\ref{dimensionless2}).

Following steps similar to those performed in~\cite{Mir,Mir2,Parisio}, we
define Fourier expansions for the velocity potentials 
\begin{equation}
\phi_{j}=\sum_{n \ne 0}\phi_{jn}(t)~\exp{ \left [ i n \varphi + (-1)^{j - 1} ~
\frac{|n|}{a} z \right ]},
\end{equation}
which obey Laplace's equation and vanish as $z \rightarrow \pm \infty$. We 
express $\phi_{j}$ in terms of the perturbation
amplitudes $\zeta_n$ by considering the kinematic boundary condition
for flow in a cylindrical cell. As in the flat cell case, the kinematic 
condition 
${\bf n} \cdot {\bf v}_{1}~|_{z=\zeta}={\bf n} \cdot {\bf v}_{2}~|_{z=\zeta}$ 
refers
to the continuity of the normal velocity across the fluid-fluid
interface. Substituting these relations into
Eq.~(\ref{dimensionless2}), and Fourier transforming, yields the
mode coupling equation of the Saffman-Taylor problem in a cylindrical
Hele-Shaw cell
\begin{equation}
\label{modecoupling}
\dot{\zeta}_{k}=\lambda(k)~\zeta_{k} + \sum_{k' \neq 
0}G(k,k')~\dot{\zeta}_{k'}\zeta_{k - k'} +  {\cal O} (\zeta_{k}^{3}),
\end{equation}
conveniently written in terms of the characteristic wave number $k=n/a$. 
The overdot denotes total time derivative, and
\begin{equation}
\label{lambda}
\lambda(k)=F(b/a)~|k|[U  - k^{2}]
\end{equation}
is the dimensionless linear growth rate. The function
\begin{equation}
\label{thirdterms}
G(k,k')=A |k| \left[ 1 - sgn(kk') \right]
\end{equation}
is the second order mode-coupling term, where the $sgn$ function equals $\pm 
1$ according to the sign of its argument.

Based on mode coupling Eq.~(\ref{modecoupling}) we now discuss some noteworthy 
features of both linear and weakly nonlinear regimes. 
Start with the linear growth rate 
~(\ref{lambda}): it is written as the product of the correction factor 
$F(b/a)$ by $\lambda(k)_{rect}=|k|[U  - k^{2}]$, which incidentally, is the 
linear growth rate for the flat, rectangular case~\cite{Mir}. This last 
observation indicates that, for fixed gap thickness $b$, there is a slight 
decrease of linear growth for increasingly larger radius of curvature $a$. 
The linear solution to Eq.~(\ref{modecoupling}) is 
purely exponential $\zeta_{k}^{lin}(t)=\zeta_{k}(0)\exp[\lambda(k)t]$, and 
introduces the correction factor $F(b/a)$ into the linear, rectangular 
solution~\cite{Mir}. At the linear level, this correction becomes more and 
more important as time progresses and when $b/a \rightarrow 1$.

From Eq.~(\ref{lambda}) we can extract two relevant
parameters: (i) the critical wave number $k_{c}=1/\sqrt{U}$ [defined by setting
$\lambda(k)=0$], beyond which all modes are linearly stable;
and (ii) the fastest growing mode $k^{*}=k_{c}/\sqrt{3}$ which maximizes 
$\lambda(k)$, and dominates the initial dynamics of the interface. 
Note that $k_{c}$ and $k^{*}$ show no dependence on the 
radius of curvature $a$. This behavior is illustrated in Fig. 2 that 
depicts the linear growth rate~(\ref{lambda}) as a function of $k$, 
in the flat cell limit (dashed curve) and for a cylindrical cell with $a 
\approx b$ (solid curve). Note that for both cases the peak location 
and width of the band of unstable modes remain unchanged, independently 
of the value of $a$. Differences in behavior between flow in cylindrical and 
flat cells are more pronounced around $k^{*}$. These facts can be interpreted 
as follows: since $k=n/a$, if one increases the cell's radius of curvature 
$a$, the number of fingers $n$ at the two-fluid interface is also increased in 
order to keep the fastest growing $k$ constant.

Now we turn our attention to the weakly nonlinear flow stage. We begin by 
discussing finger competition dynamics. It is well known that 
the viscosity contrast $A$ has a crucial role in determining interfacial 
behavior for flow in flat rectangular cells~\cite{Try1,Zhao,DiF1,DiF2,Mir}. 
For low viscosity contrast ($A \approx 0$) the interface is nearly up-down 
symmetric, and increasingly larger asymmetry is observed for larger values of 
$A$ ($A \approx \pm 1$). Consequently, $A$ has great influence on the dynamics 
of finger competition and pattern selection. In cylindrical cells, in addition 
to the parameter $A$, it is of interest to examine how 
finger competition dynamics is affected by radius of curvature $a$ (or 
correspondingly, by the correction factor $F(b/a)$).

In order to investigate finger competition we consider the influence of a
fundamental mode, on the growth of its sub-harmonic. To do that we
rewrite the net perturbation~(\ref{z}) in terms of
cosine and sine modes, where the cosine $a_{k}=\zeta_{k} + \zeta_{-k}$ 
and sine $b_{k}=i \left ( \zeta_{k} - \zeta_{-k} \right )$ amplitudes 
are real-valued. Then, for consistent second order expressions, we 
replace the time derivative terms $\dot{a}_{k}$ and
$\dot{b}_{k}$ on the right hand side of Eq.~(\ref{modecoupling}) by 
$\lambda(k)~a_{k}$ and
$\lambda(k)~b_{k}$, respectively. We 
consider a dominant
fundamental of wave number $k_f=k^{\ast}$, and a sub-harmonic of wave number 
$k_s=k_{f}/2$ and
relatively weaker amplitude. Without loss of generality we may 
take $a_{k_f} > 0$ and $b_{k_f}=0$. Under these circumstances, we obtain the
following equations of motion for sine and cosine sub-harmonic
\begin{equation}
\label{illustration1}
\dot{a}_{k_s}=\lambda(k_s)[1 + A k_{s}a_{k_f}]~{a}_{k_s},
\end{equation}
\begin{equation}
\label{illustration2}
\dot{b}_{k_s}=\lambda(k_s)[1 - A k_{s}a_{k_f}]~{b}_{k_s}.
\end{equation}
Note that if $A>0$ and $a_{k_f}>0$, the fundamental accelerates the
growth of the sub-harmonic cosine and inhibits the sub-harmonic sine mode.
This causes increased variability among the lengths of fingers of 
less viscous fluid 1 penetrating more viscous fluid 2. This effect describes 
finger competition. As was in flat cells, interface asymmetry and 
finger competition are enhanced to a degree proportional to $A$. 

Although the second order term $G(k,k')$ (see Eq.~(\ref{thirdterms})) has no 
explicit dependence on $a$, the coupling between modes $k_{f}$ and $k_{s}$ 
makes the interface dynamics sensitive to the background cylindrical geometry. 
From equations~(\ref{illustration1}) and~(\ref{illustration2}) we see that for 
a given $A$, the degree of competition and up-down asymmetry depend on the 
overall multiplicative term $\lambda(k_s)$, which in turn is proportional to 
the correction factor $F(b/a)$. To illustrate the combined influence of $A$ 
and $a$ on finger competition, we plot in Fig. 3 the effective growth rate 
$\lambda_{eff}=\lambda(k_s)[1 + A k_{s}a_{k_f}]$, taken from 
Eq.~(\ref{illustration1}), as a function of viscosity contrast $A$. In Fig. 3 
the dashed line corresponds to the flat cell limit, while the solid line 
expresses behavior for a cylindrical cell with $a \approx b$. We see from Fig. 
3 that the discrepancies between flow in cylindrical and flat cells are more 
noticeable for increasingly larger values of {\it both} $A$ and $b/a$. 
Therefore, 
with respect to finger competition, the flow will be more stable with the 
increase of the cylindrical cell radius $a$. Geometrically speaking, we can 
say that the mean curvature $H=(1/2a)$ of the cylindrical cell intensifies the 
competition among fingers in comparison with that of rectangular, planar flow.

We conclude this section by briefly discussing the possibility of occurrence 
of finger tip-splitting in cylindrical cells. Tip-splitting is related to the 
influence of a fundamental mode $k_{f}$ on the growth of its harmonic 
$k_{h}=2k_{f}$~\cite{Mir}. By rewriting Eq.~(\ref{modecoupling}) in terms of 
sine and cosine modes, and considering the coupling between $k_{f}$ and 
$k_{h}$, we verify that the harmonic mode cannot be influenced by the 
fundamental. Therefore, at second order, there is no tendency for the fingers 
to split in cylindrical cells. Tip-splitting is absent for any value of the 
cell's radius of curvature $a$, including the flat cell limit $a \rightarrow 
\infty$. This fact may be interpreted in geometric terms as follows: unlike 
finger competition behavior, which depends on, and varies with the cylinder's 
{\it mean} curvature $H$, finger tip-splitting is controlled by {\it Gaussian} 
curvature~\cite{Parisio} which is zero for a cylinder. In this sence, the 
absence of tip-splitting in cylindrical cells was expected.

\section{Conclusions and perspectives}
\label{final}
In this paper we investigated viscous flow in cylindrical Hele-Shaw cells 
analytically. The study of flow in such geometry requires modification of 
Darcy's law equation. A generalized version of Darcy's law was derived 
from first principles. It introduces pertinent corrections to usual Darcy's 
law in flat, rectangular geometry. We used a mode-coupling 
approach to examine the fluid-fluid interface evolution. Arbitrary viscosity 
contrast $A$ and cell radius $a$ have been considered. We deduced the 
following general results: on the linear level, there is an inhibition of 
growth for increasingly larger radius of curvature $a$; and, for sligthly 
nonlinear stages, we found that finger competition and interface asymmetry are 
enhanced for flow in cylindrical cells. In addition, we explained the absence 
of tip-splitting events in cylindrical cells.

The study of geometry-related corrections $F(b/a)$ for the case 
$b/a>1$ could be interesting to study in the future. Within this limit, 
and following the lines of recent work by Ruyer-Quil~\cite{Quil} and by 
Meignin {\it et al.}~\cite{Meignin}, it would be of interest to study inertial 
corrections to the generalized Darcy's law~(\ref{ave}), and examine gap size 
effects for the Saffman-Taylor instability in cylindrical cells. Experimental 
study in this direction could use, and take advantage of already existing, 
very good cylindrical Taylor-Couette cells. In addition, a thorough 
investigation of fully nonlinear flow stages in cylindrical cells, through 
extensive computer simulations, may reveal additional corrections and new 
dynamic behavior. Such numerical studies could provide a more meaningful 
confrontation between experiment and theory in cylindrical cells.

\vspace{0.5 cm}
\begin{center}
{\bf ACKNOWLEDGMENTS}
\end {center}
\noindent
I thank CNPq and FINEP (through its PRONEX Program) for financial support. 
Helpful discussions with Fernando Parisio and Claudio Furtado are gratefully 
acknowledged.

\pagebreak
\noindent
\centerline{{\large {FIGURE CAPTIONS}}}
\vskip 0.5 in
\noindent
{FIG. 1:} Schematic configuration of viscous flow in a cylindrical Hele-Shaw 
cell. The dashed curve represents the unperturbed interface $z=0$ and the 
solid undulated curve depicts the perturbed interface $z=\zeta(\varphi,t)$. 
All other relevant quantities are defined in the text.
\vskip 0.25 in
\noindent

\vskip 0.5 in
\noindent
{FIG. 2:} Variation of the dimensionless growth rate~(\ref{lambda}) 
as a function of $k$, for $U=1$: in the flat cell limit $b/a \rightarrow 0$ 
(dashed curve), and in the cylindrical case $b/a \rightarrow 1$ (solid curve).
Note that for both situations $k_{c}=1$ and $k^{*}=1/\sqrt{3}$.
\vskip 0.25 in
\noindent

\vskip 0.5 in
\noindent
{FIG. 3:} Variation of the effective growth rate  
$\lambda_{eff}=\lambda(k_s)[1 + A k_{s}a_{k_f}]$, from 
Eq.~(\ref{illustration1}), as a function of $A$, for $U=1$ and $a_{k_f}=1$: in 
the flat cell limit $b/a \rightarrow 0$ (dashed line), and in the cylindrical 
case $b/a \rightarrow 1$ (solid line). For a given value of viscosity contrast 
$A$, finger competition increases for larger values of $b/a$.
\vskip 0.25 in
\noindent


\begin{thebibliography}{99}

\bibitem{Saf}P. G. Saffman and G. I. Taylor, Proc. R. Soc. London
Ser. A {\bf 245}, 312 (1958).

\bibitem{Rev}For review articles on this subject, see D. Bensimon,
L. P. Kadanoff, S. Liang, B. I. Shraiman, and C. Tang, Rev. Mod.
Phys. {\bf 58}, 977 (1986); G. Homsy, Ann. Rev. Fluid Mech.
{\bf 19}, 271 (1987); K. V. McCloud and J. V. Maher, Phys. 
Rep. {\bf 260}, 139 (1995).

\bibitem{Tho}B. W. Thompson, J. Fluid Mech. {\bf 31}, 379 (1968).

\bibitem{Try1}G. Tryggvason and H. Aref, J. Fluid Mech. {\bf 136}, 1 (1983).

\bibitem{Zhao}H. Zhao and J. V. Maher, Phys. Rev. A {\bf 42}, 5894 (1990).

\bibitem{Cas}F. X. Magdaleno and J. Casademunt, Phys. Rev. E {\bf 63}, 043102 
(2001), and references therein.

\bibitem{DiF1}M. W. DiFrancesco and J. V. Maher, Phys. Rev. A {\bf 39}, 4709 
(1989). 

\bibitem{DiF2}M. W. DiFrancesco and J. V. Maher, Phys. Rev. A {\bf 40}, 295 
(1989).

\bibitem{Park}C. W. Park and G. M. Homsy, J. Fluid Mech. {\bf 139}, 291 (1984).


\bibitem{Maher}J. V. Maher, Phys. Rev. Lett. {\bf 54}, 1498 
(1985).

\bibitem{Max}T. Maxworthy, J. Fluid Mech. {\bf 177}, 207 (1987).


\bibitem{Mir}J. A. Miranda and M. Widom, Int. J. Mod. Phys. B 
{\bf 12}, 931 (1998).

\bibitem{Mir2}J. A. Miranda, F. Parisio, F. Moraes, and M. Widom, Phys. Rev. E 
{\bf 63}, 016311 (2001). 


\bibitem{Parisio}F. Parisio, F. Moraes, J. A. Miranda, and M. Widom, Phys. 
Rev. E {\bf 63}, 036307 (2001). 

\bibitem{Quil}C. Ruyer-Quil, C. R. Acad. Sci. Ser. IIb: Mec., Phys., Chim., 
Astron. {\bf 329}, 1 (2001).

\bibitem{Meignin}L. Meignin, P. Ern, P. Gondret, and M. Rabaud, Phys. Rev. E 
{\bf 64}, 026308 (2001).





\end{thebibliography}
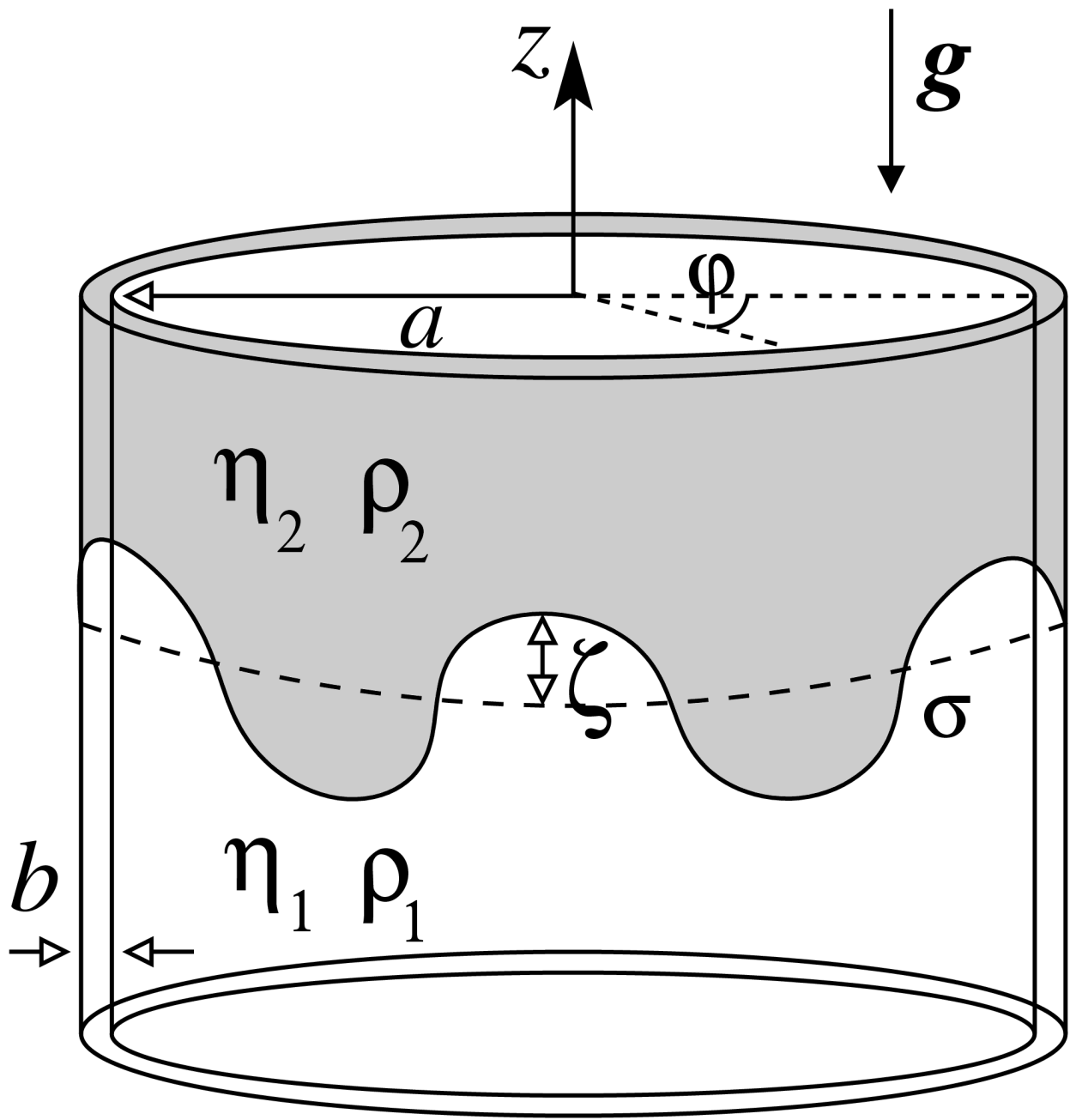
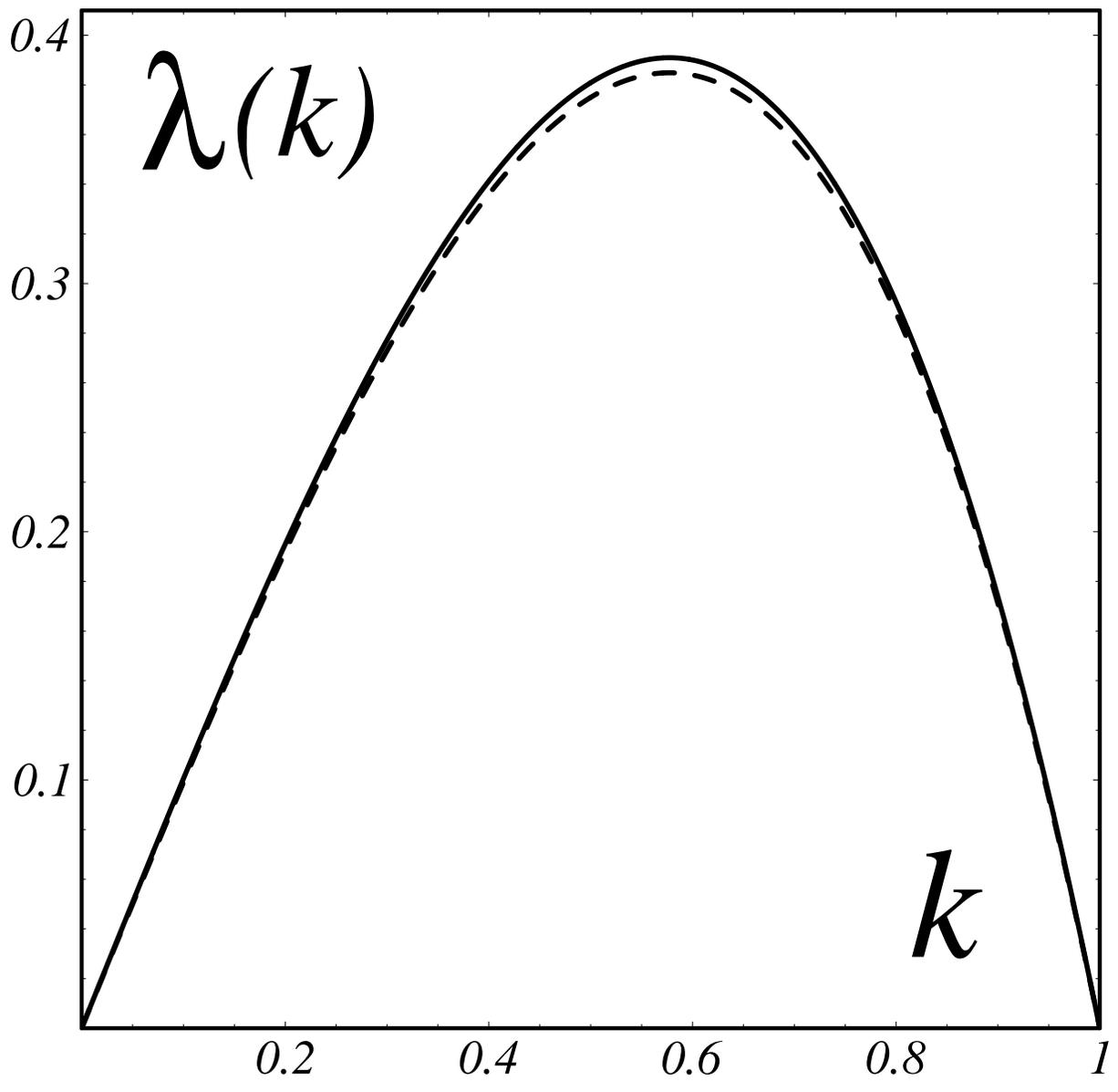
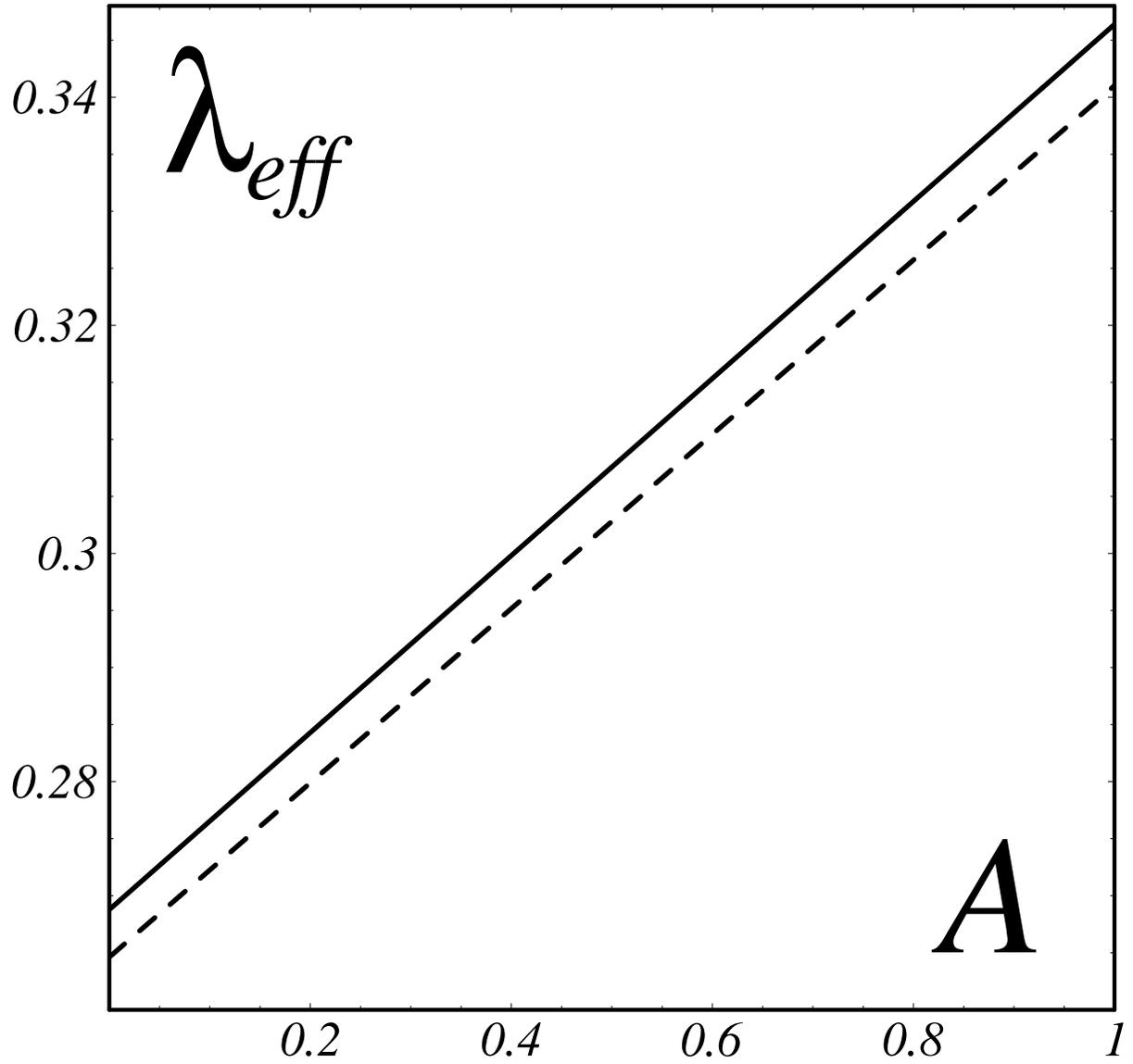
\end{document}